\documentstyle[12pt,epsf,aps]{revtex}

\begin{document}
\newcommand{\beq}{\begin{equation}}
\newcommand{\eeq}{\end{equation}}
\newcommand{\beqa}{\begin{eqnarray}}
\newcommand{\eeqa}{\end{eqnarray}}
\draft
\title{Off shell behaviour of the in medium nucleon-nucleon 
cross section}
\author{C. Fuchs$^1$, Amand Faessler$^1$, M. El-Shabshiry$^2$}
\address{
$^1$Institut f\"ur Theoretische Physik, Universit\"at T\"ubingen, 
T\"ubingen, Germany \\
$^2$Physics Department, Faculty of Science, Ain Shams University, 
Cairo, Egypt
}
\maketitle

\begin{abstract}
The properties of nucleon-nucleon scattering inside dense 
nuclear matter are investigated. We use the relativistic 
Brueckner-Hartree-Fock model to determine on-shell and 
half off-shell in-medium transition amplitudes and cross sections. 
At finite densities the on-shell cross sections are generally 
suppressed. This reduction is, however, less pronounced than 
found in previous works. 
In the case that the outgoing momenta are allowed to be off energy 
shell the amplitudes show a strong variation with momentum. 
This description allows to determine in-medium cross sections 
beyond the quasi-particle approximation accounting thereby for 
the finite width which nucleons acquire in the dense nuclear medium. 
For reasonable choices of the in-medium nuclear spectral width, 
i.e. $\Gamma\leq 40$ MeV, the resulting total cross sections are, 
however, reduced by not more than about 25\% compared to the 
on-shell values. Off-shell effect are generally more pronounced at 
large nuclear matter densities.
\end{abstract}

\section{Introduction}
One major topic of modern nuclear physics is the investigation of 
hadron properties inside a dense hadronic environment which 
exists, e.g. in the interior of neutron stars or is transiently created 
in energetic heavy ion collisions. In the latter case, 
the theoretical framework to 
describe the time evolution of heavy ion reactions is provided by 
kinetic transport theory. Starting from the quantum theory of 
strongly interacting Fermi systems, formulated within the framework of the 
Dyson-Schwinger hierarchy of non-equilibrium many-body Green 
functions \cite{ms59} one can derive semi-classical transport 
equations of a Boltzmann-Uehling-Uhlenbeck (BUU) type 
\cite{glw80,btm90}. These 
transport equations describe successfully 
the time evolution of a non-equilibrated strongly 
interaction hadron gas. To mention 
only the essential steps of such a derivation there are: the truncation 
of the many-body hierarchy at the two-body level, a Wigner 
transformation of the density matrices with subsequent gradient 
expansion up to first order in $\hbar$ and the usage of the quasi-particle 
approximation (QPA) which neglects the finite decay width of the 
particles. The resulting BUU equation consists of two parts, 
a drift term which propagates 
the particles dressed by the surrounding medium in a self-consistent 
mean field, and the collisions term responsible for binary 
nucleon-nucleon scattering. In a consistent treatment both ingredients, 
namely the mean field and the binary cross sections should be treated 
on the same footing which means to base both of them on the same 
effective interaction. Unfortunately, in most applications to 
heavy ion collisions this is not 
done. The self consistent mean field accounts for medium effects 
by its density dependence. For the cross section, on the other 
hand side, the free (vacuum) expressions are widely 
used in transport calculations. It has, however, been 
noticed that in particular at incident 
energies below the particle production thresholds, medium modifications 
of the cross sections can play an important role for the reaction 
dynamics in heavy ion collisions \cite{jae92,gmat2,zheng}.

The relativistic (Dirac-) Brueckner approach 
\cite{hs87,thm87,nupp89,bm90,sehn97,fuchs98,boelting99} provides a 
powerful tool to achieve such a consistent description. 
Starting from free nucleon-nucleon 
interactions, given in its modern form by one-boson-exchange 
potentials \cite{bonn} one treats the two-body correlations 
in dense nuclear matter in the ladder approximation of the 
Bethe-Salpeter equation. As a result the nuclear matter saturation 
properties are quite well described. This is achieved 
without the adjustment of additional parameters, as e.g. done 
in relativistic mean field 
models \cite{sw86}. On the level of the T-matrix approximation 
both ingredients for the BUU equation follow from the on-shell 
in-medium T-matrix (or G-matrix). The mean field is determined by the 
real part $Re T$ of the T-matrix whereas the cross section 
$\sigma \simeq |T|^2$ is 
connected to the imaginary part $Im T$ via an optical theorem \cite{btm90}. 
Medium modifications arise due to the dressing of the quasi-particles 
and the existence of the Pauli operator which prevents the 
scattering of intermediate states in the Bethe-Salpeter equation 
(not final states) into occupied 
phase space areas. Both aspects are most pronounced at high densities 
and/or low momenta and lead to a suppression of the in-medium 
cross section compared to the free one. 

There have already several studies been devoted to the 
in-medium NN scattering problem. The T\"ubingen group \cite{jae92} and 
later on the Rostock group \cite{schmidt90,alm94} derived 
in-medium cross sections within the non-relativistic Brueckner 
approach, in the latter case also at finite 
temperature. Relativistic calculations were performed in 
\cite{thm87b,malfliet88,cross}. The most complete study of 
in-medium NN scattering within the Dirac-Brueckner approach 
was probably done by Li and Machleidt \cite{lima93} who used the 
BONN potentials as bare interaction. Unfortunately the different 
approaches have led to partially contradictory results, 
in particular between relativistic \cite{thm87b,lima93} and the 
non-relativistic calculations \cite{jae92}. 
Therefore, in the first part of the 
paper we revisit the problem of on-shell scattering. The results 
are obtained with the BONN A potential. In large parts we find a good 
agreement with the previous investigations of Li and Machleidt \cite{lima93}. 
However, their treatment \cite{lima93} seems to overestimate the 
suppression of the in-medium cross section at low energies compared to the 
vacuum case. 

The second part of the paper is devoted to an additional aspect: 
Kinetic transport equations are essentially based on the quasi-particle 
approximation (QPA) which puts the nucleons on the mass-shell. 
The same holds in the medium for the dressed quasi-particles. 
If the imaginary part of the nucleonic self-energy is 
negligible ($Im\Sigma \ll Re\Sigma$) the quasi-particle approximation 
as the zero-width limit ($\Gamma \propto Im\Sigma$) 
for the nucleon spectral function appears 
to be justified \cite{btm90}. However, it is well known that 
the spectral widths of hadrons change in the medium. It has further 
been pointed out that in the medium also ``stable'' particles can obtain a 
non-zero width, depending on their collision 
rates (collisional broadening), see e.g. \cite{knoll98}. Following 
the work of Botermans and Malfliet \cite{btm90} there have been 
several attempts to derive transport equations for non-equilibrated 
Fermi systems beyond the quasi-particle approximation, see  
\cite{knoll98,henning95,fauser95,morawetz99} and references therein. 
However, due to the 
complications which arise giving up the QPA these transport 
equations were never used in practical applications to 
heavy ion reactions but stayed more or less on the level of 
academical considerations. Just very recently there have been 
successful attempts to formulate generalised transport equations which 
can be handled in applications using testparticle methods 
\cite{leupold,cassing}. As the basic feature of these approaches  
energy and momentum of the testparticles are no more related 
by the mass-shell condition but according to their 
spectral distributions. Thus particles are propagated and 
also scattered off mass-shell. Hence, the knowledge of 
off-shell scattering amplitudes becomes necessary. 
However, the behaviour of such amplitudes is presently 
unknown to large extent. The off-shell structure of the 
scattering amplitude determines in this context also the magnitude 
of non-local corrections  to the Boltzmann equation which can be 
translated into non-local time and momentum shifts in the binary 
scattering process \cite{morawetz}.

Here again the relativistic Brueckner approach provides the 
natural tool to determine in-medium off-shell scattering amplitudes 
in the ladder approximation. In the present 
work we extend the on-shell NN scattering 
to the half-off-shell case where the incoming particles are 
still on their mass shells but the final states are generally 
off-shell. These matrix elements provide valuable 
information for future transport investigations beyond the QPA. 
We investigate the off-shell structure of the in-medium T-matrix, 
respectively the transition amplitudes $|T|^2$, and the resulting 
cross sections over a wide range of nuclear matter densities.

The paper is now organised as follows: First (Sec. II) we briefly 
sketch the basic features of the relativistic Brueckner approach. 
In Sec. III the on-shell scattering problem in the medium is 
discussed. Neutron-neutron and proton-proton channels are considered 
separately and total and differential cross section are given. We 
also compare with results of other groups, mainly those of 
Li and Machleidt \cite{lima93}. In Sec. IV we turn to the half 
off-shell case and discuss the structure of transition amplitudes 
and cross sections beyond the quasi-particle approximation and 
summarise in Sec. V. 
\section{Relativistic Brueckner approach}
In the relativistic Brueckner approach the Bethe-Salpeter 
equation is reduced to a three dimensional integral equation of
the Lippmann-Schwinger type, the so called Thompson equation \cite{15}.
The Thompson propagator projects thereby the intermediate nucleons 
onto positive energy states and restricts the exchanged energy
transfer by $\delta(k^0)$ to zero. 
The Thompson equation is most easily solved in the two-nucleon c.m.-frame 
\begin{equation}                
T({\bf p},{\bf q},{\bf P})|_{\rm c.m.} = V({\bf p},{\bf q}) 
+ \int \frac{d^3 {\bf k}}{(2\pi)^3}  V({\bf p},{\bf k}) 
\frac{M^{\ast 2}}{ E^{\ast 2}({\bf k}) }
\frac{Q({\bf k},{\bf P})}{ 2E^{\ast}({\bf q}) - 2E^{\ast}({\bf k}) 
+ i\epsilon } T({\bf k},{\bf q},{\bf P})      
\label{tom1}
\end{equation}
where ${\bf q}=({\bf q}_1-{\bf q}_2)/2
=({\bf q}^{\ast}_1-{\bf q}^{\ast}_2)/2$ 
is the relative momentum of the initial states and similar ${\bf p,k}$ 
are the relative momenta of the final and intermediate states, respectively.
${\bf P}= ({\bf q}_1 + {\bf q}_2)$ is the c.m. momentum.
The starting energy in Eq. (\ref{tom1}) is fixed to 
$ \sqrt{s^{\ast}} = 2E^{\ast}({\bf q})$.
Sandwiching the one-boson-exchange potential $V$ (\ref{tom1}) between 
in-medium spinors (\ref{spinor}) the matrix elements acquire 
a density dependence which is absent in 
non-relativistic treatments and which is believed to be the major 
reason for the much improved description of the nuclear 
saturation properties \cite{btm90} in the relativistic theory. 
The Pauli operator $Q$ prevents intermediate states 
from scattering into forbidden phase space areas. 

Inside the medium the particles are dressed which leads to 
effective masses and the kinetic momentum
        \begin{equation}                                
        M^{\ast}(k) = M + {\rm Re}\Sigma_s(k) \, , \qquad 
        k^{\ast\mu} = k^{\mu} + {\rm Re}\Sigma^{\mu}(k) .
\label{effmass}
        \end{equation}

Re and Im denote real and imaginary part since  (above the Fermi 
surface) the self-energy is generally complex.
Here we adopt the {\it quasiparticle approximation}, i.e.,
the Im $[\Sigma]$ will be neglected in Eq. (\ref{tom1}). 
This means that the decay width of the dressed nucleon state $k$ 
to another state $k^\prime$ is set 
equal to zero, resulting in an infinite lifetime 
of this 'quasiparticle' state. 
Furthermore, the explicit momentum dependence of the 
self-energy which enters via a term ${\bf k} \Sigma_v$ 
proportional to the spatial component  $\Sigma_v$ of the vector 
self-energy is small and  
can be dealt by introducing the reduced kinetic
momentum $\tilde k^{\ast\mu} = k^{\ast\mu} / (1+\Sigma_v)$ and the reduced 
effective mass $\tilde M^{\ast} = M^{\ast} / (1+\Sigma_v)$ 
\cite{hs87}. Thus, the nucleons are given by plane waves which 
fulfil a quasi-free Dirac equation 
\begin{equation}                     
        \left[ \gamma_{\mu} \tilde k^{\ast\mu} - \tilde M^{\ast} \right] 
        u_{\lambda}(k) = 0 
\quad .
 \label{dirac}
\end{equation}
Using the normalisation of Ref. \cite{thm87} 
the self-consistent positive-energy spinors of helicity $\lambda$ 
are defined as
\begin{equation}                   
  u_{\lambda}(k) = \sqrt{\frac{\tilde E^{\ast}(k) + \tilde M^{\ast}}
                            { 2 \tilde M^{\ast}}                  }
  \left( \begin{array}{c}  1  \\
      \frac{2 \lambda |{\bf k}| }{\tilde E^{\ast}(k) + \tilde M^{\ast}} 
         \end{array} 
  \right) \chi_{\lambda}
\label{spinor}
\end{equation}
with $\chi_{\lambda}$ being a Pauli spinor. The Dirac spinors depend on 
the effective mass and thus on the nuclear density. In the 
Thompson equation (\ref{tom1}) and in the discussion below we deal 
with the rescaled quantities ${\tilde M^{\ast}}$ and ${\tilde k^{\ast}}$ 
but will omit this in the notation further on. 

To summarise the kinematics of the Thompson equation:\\
1. The initial states are on-shell, i.e. 
$q_\mu = \{ E^{\ast}({\bf q}), \pm {\bf q}\}$ with 
$ E^{\ast}({\bf q}) = \sqrt{M^{\ast 2} +  {\bf q}^2} = \frac{1}{2}\sqrt{s^*}$. 
The final states fulfil energy-momentum conservation 
$p_\mu = \{ \frac{1}{2}\sqrt{s^*}, \pm {\bf p}\}$ and are off-shell 
as soon as $|{\bf p}| \not= |{\bf q}|$. \\
2. The determination of the off-shell matrix elements is perturbative 
in the sense that the quasiparticle approximation is applied to the 
Thompson equation, although $T$ is generally complex for incident momenta 
above the Fermi surface which leads to a non-vanishing imaginary part 
of the self-energy $Im\Sigma$ 
and correspondingly, an imaginary optical potential \cite{thm87,sehn97}.

To determine the scalar $\Sigma_s$ and vector components $\Sigma_0$ and 
$\Sigma_v$ of the self-energy is a subtle problem. Here on-shell 
ambiguities arise form the projection onto positive energy states 
when the T-matrix is decomposed into Lorentz invariant 
amplitudes. This problem has been known for a long time \cite{thm87,nupp89} 
and is still not completely resolved. In \cite{fuchs98} we discussed 
the failure of previously used recipes \cite{thm87,sehn97} which 
lead to spurious contributions in the self-energy from the coupling 
to negative energy states, in particular spurious contributions from 
a pseudo-scalar one-pion-exchange which are not completely replaced by 
a pseudo-vector coupling. In Ref. \cite{boelting99} 
this problem was extensively discussed and a method to minimise 
the on-shell ambiguities was proposed. Here we used the scheme 
of \cite{boelting99} where the Born term $V$ and the remaining 
ladder kernel of the Thompson equation are treated separately. 
Thus we account properly for the pseudo-vector structure of 
the Born contributions $V_{\pi,\eta}$ from $\pi$ and $\eta$ exchange 
contributions when the projection of the full $T$-matrix 
onto covariant amplitudes is performed. The remaining 
ladder kernel thereby is treated as pseudo-scalar.

To solve the Thompson equation (\ref{tom1}) 
in the c.m.-system we apply standard 
techniques which are in detail described by Erkelenz \cite{erkelenz}. 
After a partial wave 
projection onto the $|JMLS>$-states the integral reduces to a one-dimensional 
integral over the relative momentum $|{\bf k}|$ and Eq. (\ref{tom1}) decouples
into three subsystems of integral equations for the uncoupled spin singlet,
the uncoupled spin triplet and the coupled triplet states. 
Due to the antisymmetry of the two-fermion states we can restore the total
isospin (I=0,1) of the two-nucleon system with the help of the 
selection rule $(-)^{L+S+{\rm I}}=-1$ which means that 
matrix elements are already antisymmetrized.

The Pauli operator $Q$ is replaced by an angle averaged Pauli operator
$\overline Q$. For non-vanishing c.m. momenta the Fermi sphere 
is in the two-nucleon c.m.-frame deformed to a Fermi 
ellipsoid for which $\overline Q$ has to be evaluated \cite{hs87,thm87}.
We are solving the integral equations by the 
matrix inversion techniques of Haftel and Tabakin \cite{11}. 
Real and imaginary parts of the T-matrix are calculated separately 
by the principal-value treatment given by Trefz {\it et al.} \cite{trefz}. 
From there it is seen that $Im T$ is essentially proportional to 
the angle averaged Pauli operator and thus it is strongly reduced 
for momenta below the Fermi surface due to Pauli blocking. 
Then positive-energy helicity T-matrix elements are constructed from the 
$|JMLS>$-scheme as described in \cite{bm90,erkelenz}. From general symmetries
it follows that for each total angular momentum $J$ 
only six of the sixteen helicity matrix elements are 
independent
\beqa
T_1 = < + + | T^J (p,q) | + + > ~&,&~
T_2 = < + + | T^J (p,q) | - - > \nonumber\\
T_3 = < + - | T^J (p,q) | + - > ~&,&~
T_4 = < + - | T^J (p,q) | - + > \nonumber\\
T_5 = < + + | T^J (p,q) | + - > ~&,&~
T_6 = < + - | T^J (p,q) | + + > 
\label{heli1}
\eeqa
which in the on-shell case $|{\bf p}| = |{\bf q}|$ further reduce 
to five independent matrix elements since then $T_5 = T_6$. 
From the six independent amplitudes in the $|JMLS>$-representation
the six independent partial wave amplitudes (\ref{heli1}) 
in the helicity representation
(for I=0,1 and real and imaginary part separately) are obtained
as described in Ref. \cite{erkelenz}.
Summation over $J$ yields the full helicity matrix elements (\ref{heli1})
\begin{equation}        
\sum_J \left[ \frac{2J+1}{4\pi} \right]
d^J_{\lambda \lambda^{\prime} }(\theta)
< \lambda^{\prime}_1 \lambda^{\prime}_2 |T^J (p,q)| \lambda_1 \lambda_2>
= <{\bf p} \lambda^{\prime}_1 \lambda^{\prime}_2 | {\hat T}
| {\bf q} \lambda_1 \lambda_2 > .
\label{heli2}
\end{equation}
Here $\theta$ is the scattering angle between ${\bf q}$ and ${\bf p}$ and 
$\lambda = \lambda_1 - \lambda_2, \lambda' = \lambda_1' - \lambda_2'$.
The reduced rotation matrices $d^J_{\lambda \lambda^{\prime} }(\theta)$ 
are those defined by Rose \cite{16}.
The matrix elements on the left hand side of Eq. (\ref{heli2}) are 
independent of the third component of the isospin ${\rm I}_3$ and 
depend only on the absolute values $p,q$ of the momenta. 
\section{On-shell scattering}
The on-shell nucleon-nucleon cross section can be directly determined 
from the T-matrix amplitudes. In this case the extension to off-shell 
scattering is straightforward. Another possibility is to 
determine the on-shell phase shifts \cite{lima93,erkelenz}. Doing so, 
an extension to the off-shell case is, however, unclear. 
Furthermore, the definition 
of the vacuum phase shifts, see e.g. refs. \cite{bm90,erkelenz}, 
has to be modified inside 
the medium to account for the modified unitarity relations. To be 
more precise, the definition of in-medium phase shifts should include  
the Pauli operator as pointed out in \cite{schmidt90,alm94}. 
To avoid such problems we will directly determine 
the cross sections from the matrix elements. 
The squared matrix elements are given as 
\beq
| {\hat T} (p,q,\theta) |^2 = 
\sum_{i=1}^{6} \beta_i \left[ 
\left( \sum_{J} \frac{2J+1}{4\pi} 
d^J_{\lambda_i \lambda^{\prime}_i }(\theta) Re T_{i}^J (p,q)\right)^2 
+ \left( \sum_{J} \frac{2J+1}{4\pi} 
d^J_{\lambda_i \lambda^{\prime}_i }(\theta) Im T_{i}^J (p,q)\right)^2
\right]
~.
\label{tsq1}
\eeq 
The weighting factors $\beta_i =2~,~i=1..4$ and 
$\beta_5 =\beta_6 =4$ arise from the sum over all helicity 
states. Using the orthogonality relation for the rotation matrices 
\beq
\int d\cos (\theta) d^J_{\lambda \lambda^{\prime} }(\theta) 
d^{J^\prime}_{\lambda \lambda^{\prime} }(\theta) 
=  \frac{2}{2J+1} \delta_{J J^\prime}
\eeq
one obtains 
\beq
\int d\Omega | {\hat T} (p,q,\theta) |^2 
= \sum_{i=1}^{6} \beta_i \sum_{J} \frac{2J+1}{4\pi} \left[ 
\left( Re T_{i}^J (p,q)\right)^2 
+ \left( Im T_{i}^J (p,q)\right)^2 
\right]
\label{tsq2}
\eeq 
In the on-shell case $p=q$ the differential cross section 
follows from the matrix elements by the standard expression  
\beq
d\sigma = \frac{(M^*)^4}{s^* 4 \pi^2} | {\hat T} (q,q,\theta) |^2 
d\Omega~.
\label{sig1}
\eeq
\subsection{Free cross section}
The predictions of the Bonn potentials for free NN cross sections 
have in detail been discussed by Li and Machleidt \cite{lima93}. 
Throughout this work we apply the Bonn A potential \cite{bonn} and the results 
of \cite{lima93} for the vacuum case are reproduced with high accuracy 
for both, differential and total cross sections. To demonstrate 
this feature, the results from \cite{lima93} are included in 
Figs.\ref{fig7} and \ref{fig8} where the total neutron-proton 
($ T_{i}^J = 0.5 (T_{i}^{J, I=0} + T_{i}^{J, I=1})$) and 
proton-proton ($T_{i}^J = T_{i}^{J, I=1}$) cross section are shown. 
As found in \cite{lima93} the $pp$ cross section is in 
particular at low energies significantly smaller 
than the $np$ cross section. It should, however, be noted that in 
the present work as well as in ref.  \cite{lima93} the $pp$ 
cross sections are not Coulomb corrected. 
\subsection{In-medium cross section}
As already pointed out in refs. \cite{thm87b,malfliet88} a trivial 
medium modification of the cross sections arises from the 
in-medium masses entering into the kinematical term $(M^*)^4/s^*$ 
in eq. (\ref{sig1}) which is due to the normalisation of the relativistic 
spinor basis and the incoming flux. This phase space 
 factor reduces the in-medium cross 
section by the order of $(M^* /M)^2$ at small momenta.  

\begin{figure}
\begin{center}
\leavevmode
\epsfxsize = 10cm
\epsffile[100 50 460 400 ]{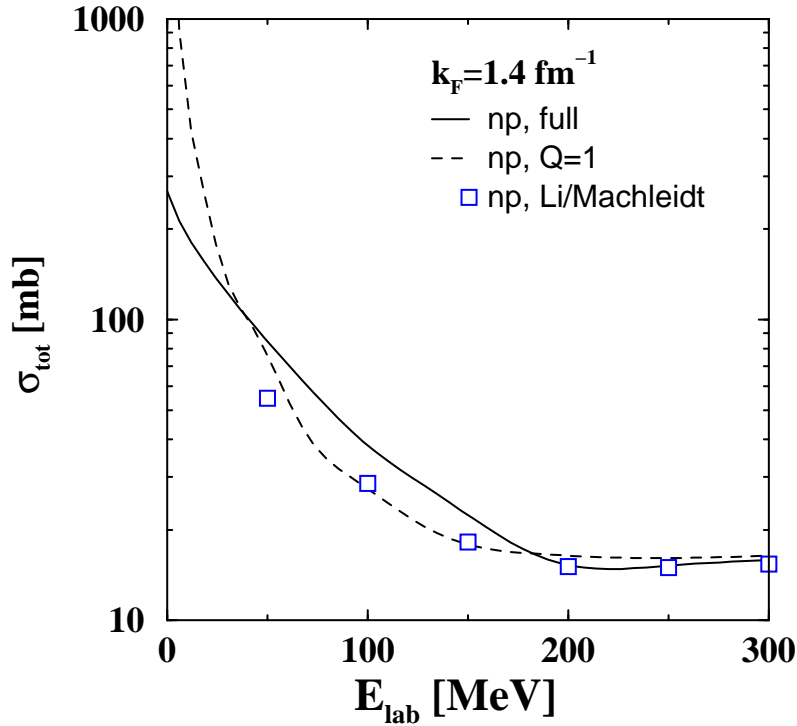}
\end{center}
\caption{Total $np$ in-medium cross section at 
$k_F=1.4~{\rm fm}^{-1}$. The result of the full calculation is 
compared to a calculation where the Pauli operator in the 
Thompson-equation for the intermediate scattering states 
 has been switched off ($Q=1$). Also the corresponding 
result of Ref. \protect\cite{lima93} 
is shown.   
}
\label{pauli.fig}
\end{figure}

Besides the fact that one deals with dressed quasi-particles 
the essential feature of the Bethe-Goldstone or 
Bethe-Salpeter equation, respectively, is the occurrence of the 
Pauli operator. This means that the vacuum relations which 
connect the phase shifts with the real reaction matrix $R$ 
are modified by the Pauli operator. In a schematic 
notation\footnote{For the clarity of the notation we suppress in 
eqs. (\ref{rmat}-\ref{reimt}) factors $\frac{M^*}{E^*}$ which 
can be absorbed into $R$ and $T$ \cite{bm90} and the $\delta$-function 
originating from the principle value treatment of the 
Blankenbecler-Sugar, respectively the Thompson propagator.} 
the in-medium reaction matrix $R$ is connected to the 
$T$-matrix by \cite{bm90,trefz}
\beq 
{\hat R} - {\hat T} = i \pi {\hat R}Q{\hat T}
\label{rmat}
\eeq
which leads to a modified optical theorem 
\beqa
| {\hat T}|^2 = \frac{ {\hat R}^2 }{1 + (\pi {\hat R}Q)^2} 
= (\pi Q)^{-1} |Im {\hat T}|
\quad . 
\label{opt}
\eeqa
With $Q=1$ the vacuum expressions are recovered. The modification 
of the optical theorem by the presence of the medium, in 
particular the appearance of the inverse Pauli operator which 
compensates at momenta below the Fermi surface for the 
vanishing $Im {\hat T}$, has been discussed in 
\cite{schmidt90,alm94,trefz}. It becomes clear 
from (\ref{opt}) that the use of the vacuum relations (with $Q=1$) 
to extract phase shifts from the in-medium reaction matrix $R$ is an 
approximation justified at low densities and/or high energies. 
In between the Pauli operator is essentially different from 
unity and appears in the denominator of eq. (\ref{opt}). Neglecting 
here the influence of the Pauli operator will lead to an 
underestimation of the corresponding cross sections. 
The real and imaginary part of the T-matrix are related to the 
reaction matrix by \cite{trefz}
\beqa
Re{\hat T} = \frac{{\hat R}}{1 + (\pi {\hat R}Q)^2}~,~
Im{\hat T} = -\frac{\pi Q{\hat R}^2}{1 + (\pi {\hat R}Q)^2}~.
\label{reimt}
\eeqa
We emphasise this point because, as will be seen in the following, 
we find the in-medium cross sections to be substantially 
less suppressed at low momenta than found by Li and 
Machleidt \cite{lima93} whereas we obtain a good agreement 
with their results at high momenta. 
The reason for the deviations can be traced back to 
the different procedures used to determine the cross sections. 
As discussed in sec. 2 the squared matrix elements (\ref{tsq1}) 
provide an unambiguous and direct method to extract the cross 
section. To determine phase shifts 
first, has to be done with caution since 
the occurrence of the Pauli operator modifies the corresponding 
phase shift relations in the medium \cite{schmidt90}. If neglected, 
as done in the approximation used in Ref. \cite{lima93}, 
this effect leads in particular at low momenta 
to an underprediction of the cross section. 
To illustrate this effect in Fig. 4 we investigate the influence 
of the Pauli operator on the in-medium $np$ cross section. For 
a fair comparison the density ($k_F=1.4~{\rm fm}^{-1}$) 
as well as the value of $M^*$ are chosen as in Ref. \cite{lima93}. 
One curve in Fig.\ref{pauli.fig} is obtained by switching off 
the Pauli operator in the Thompson equation (\ref{tom1}), 
i.e. setting ${\bar Q}\equiv 1$. It is clearly seen that the influence 
of the Pauli operator leads even to an enhancement of the 
cross section at momenta below $\simeq 180$ MeV 
compared to the ${\bar Q}\equiv 1$ case which is due to the 
occurrence of $Q$ in the denominator in Eqs. (\ref{opt},\ref{reimt}). 
As expected, at very small momenta 
the presence of the Pauli operator leads to a significant suppression 
of the cross section. One should, however, keep in mind that the 
Pauli operator acts here only on the {\it intermediate} states in the 
Thompson equation (\ref{tom1}) and not on the final states. Thus 
full Pauli blocking ${\bar Q}\equiv 0$ reduces the full T-matrix 
to its Born part ${\hat V}$. In the transport approach, on the other 
hand side, the Pauli blocking prevents also the scattering into 
occupied final states. Thus Fig. 4
refers only to the Pauli effects in the intermediate states, but  
demonstrates the importance to account properly for the Pauli 
operator in the in-medium optical theorem (\ref{opt}). 

In the following we consider the in-medium cross sections at four 
different Fermi momenta $k_F = 1.1,~1.34,~1.7$ and $1.9$ MeV which 
corresponds to densities $\rho = 0.090,~0.1625,~0.332$ and 0.4633 
fm$^{-3}$. For simplicity we denote these densities in the 
text as 0.5/1/2/3 $\rho_0$ although these values do not 
exactly correspond to multiples of $\rho_0= 0.1625~fm^{-3}$. 
Again all calculations are performed using the Bonn A 
potential. The in-medium mass $M^*$ entering into eq. 
(\ref{sig1}) has in our calculation the values 
$M^* = 766.6,~646.7,~433.6,$ and $310.1$ MeV. These values are 
slightly larger than those of ref. \cite{lima93,bm90}. Thus we 
expect also slightly larger values for the in-medium cross 
section, in particular at low momenta which are due to higher values 
for the kinematical factor $(M^*)^4/s^*$. The reason for the 
different effective masses lies in different solution 
techniques for the Thompson equation 
(\ref{tom1}). As discussed in Sec. 2, we apply a refined projection 
scheme in order to transform the $T$ matrix from 
the two-particle c.m. frame to the nuclear matter rest frame 
where the self-energy components (\ref{effmass}) are determined 
\cite{fuchs98,boelting99}. In the medium the on-shell T-matrix 
(\ref{tom1}) depends on three variables, the relative momentum 
$\bf q$ of the initial states, the scattering angle $\theta$ and 
the centre-of-mass momentum $\bf P$ of the two-particle c.m. 
frame relative to the nuclear matter rest frame. As in refs. 
\cite{lima93,schmidt90,alm94} we consider only the case where the 
two-particle c.m. frame and the nuclear matter rest frame coincide, 
i.e. ${\bf P} =0$.

\begin{figure}
\begin{center}
\leavevmode
\epsfxsize = 10cm
\epsffile[100 50 460 400 ]{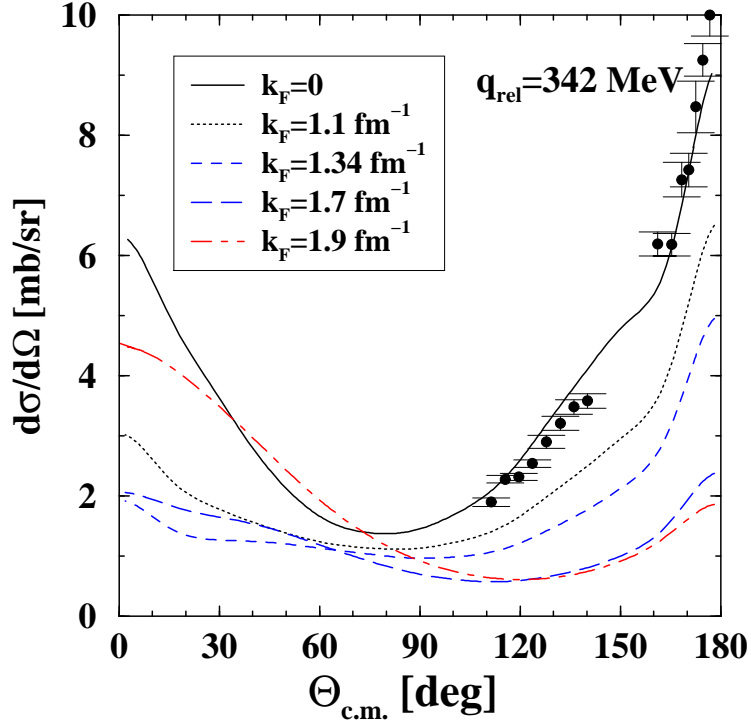}
\end{center}
\caption{Differential $np$ in-medium cross section at 
fixed relative c.m. momentum $|{\bf q}|$=342 MeV 
($\frac{2{\bf q}^2}{M}=250$ MeV) 
at various densities. Experimental data from \protect\cite{shepard74} 
for the free scattering are included.
}
\label{fig4}
\end{figure}

\begin{figure}
\begin{center}
\leavevmode
\epsfxsize = 10cm
\epsffile[100 50 460 400 ]{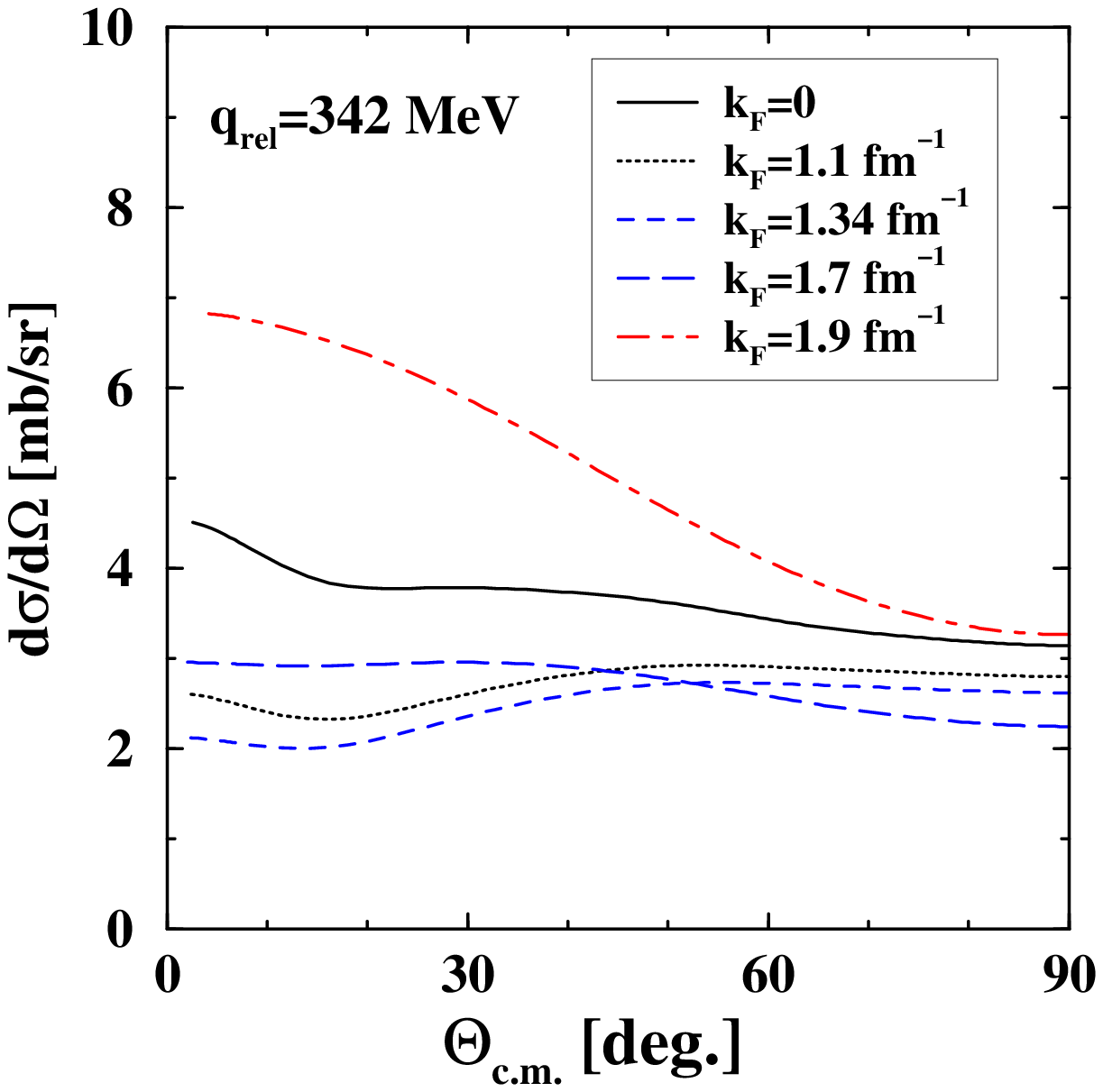}
\end{center}
\caption{Differential $pp$ in-medium cross section at 
fixed relative c.m. momentum $|{\bf q}|$=342 MeV 
($\frac{2{\bf q}^2}{M}=250$ MeV) 
at various densities.
}
\label{fig6}
\end{figure}

In Fig. \ref{fig4} the differential $np$ cross section at the 
different densities is shown at fixed relative momentum 
$|{\bf q}|$=342 MeV which in the vacuum corresponds to a laboratory energy of 
$E_{\rm lab}=\frac{2{\bf q}^2}{M}$=250 MeV. The vacuum definition 
of $E_{\rm lab}$ was used in \cite{lima93} to compare the 
cross sections at different densities. The presence of the medium tends 
to make the $np$ differential cross section more isotropic. 
At backward angles the cross section are decreasing with density. 
At forward angles the behaviour is more complicated: At moderate 
densities ($\rho=0.5/1~\rho_0$) the cross section is reduced but at high 
densities  ($\rho=2~{\rm and}~3~\rho_0$) a strong enhancement of the 
forward scattering amplitude can be observed. It is worth to 
notice that in this energy range we are in good agreement 
with the results obtained by Li 
and Machleidt \cite{lima93}. Similar results have been 
obtained at 0.5, 1 and 2$\rho_0$ and also at 2$\rho_0$ the cross section 
was found to be enhanced at forward angles compared to $\rho_0$. 
Going higher in density (3$\rho_0$) we find this effect even more 
pronounced. While the cross section stays now almost constant at 
backward angles it is strongly enhanced at forward angles. 
However, at 3$\rho_0$ the cross section is highly 
anisotropic and dominated by a $p$-wave component. Here we see a 
suppression of higher partial waves with increasing density. 
At $3\rho_0$ one needs partial waves up to 
at least $J\leq 6$ to approximate the full result ($J=12$), at $2\rho_0$ 
the partial waves $J\leq 4$ are almost sufficient 
and at $3\rho_0$ the behaviour is 
like an $s+p$-wave with $J\leq 1$. 
Fig.\ref{fig6} shows the same for $pp$ scattering. 
Again our results are in good agreement with the findings of 
ref. \cite{lima93}. In the $I=1$ channel ($pp$) the cross sections 
are generally more isotropic than in the $np$ channel. 
With rising density the cross sections 
are first decreasing ($0.5\rho_0,~\rho_0$) and then increasing. 
At $3\rho_0$ we observe a dramatic 
increase of the cross section at forward angles. 
In the $pp$ cross section the contributions of higher partial 
waves are reduced with growing density, e.g. at $\rho_0$ 
partial waves up to $J\leq 4$ reproduced the full result quite well 
whereas at $3\rho_0$ only contributions from $J\leq 3$ are relevant. 

The peculiar behaviour at $3\rho_0$ seen 
in Figs. \ref{fig4}, \ref{fig6} can 
be understood from the presence of the mean field. In 
Figs. \ref{fig4}, \ref{fig6} we investigated the density dependence 
of the cross sections at an equivalent relative c.m. momentum ${\bf q}$. 
This does, however, not correspond to equivalent energies. At finite 
density the laboratory energy 
$E_{\rm lab}(|{\bf q}|,\rho)=E(|{\bf q}|,\rho) -M$ is given 
by 
\beq
E_{\rm lab} = \frac{2{\bf q}^2}{M^*} + \Sigma_s - \Sigma_0
\label{elab}
\eeq
and is therefore strongly modified by the presence of the mean field. 
At high densities the energy scale is stretched by the 
decreasing effective mass $M^*$. This effect is responsible for 
the suppression 
of higher partial wave contribution to the differential cross 
section above $2\rho_0$ if one compares the cross sections at 
identical c.m. momenta but at essentially different incident 
energies. 

\begin{figure}
\begin{center}
\leavevmode
\epsfxsize = 10cm
\epsffile[100 50 460 400 ]{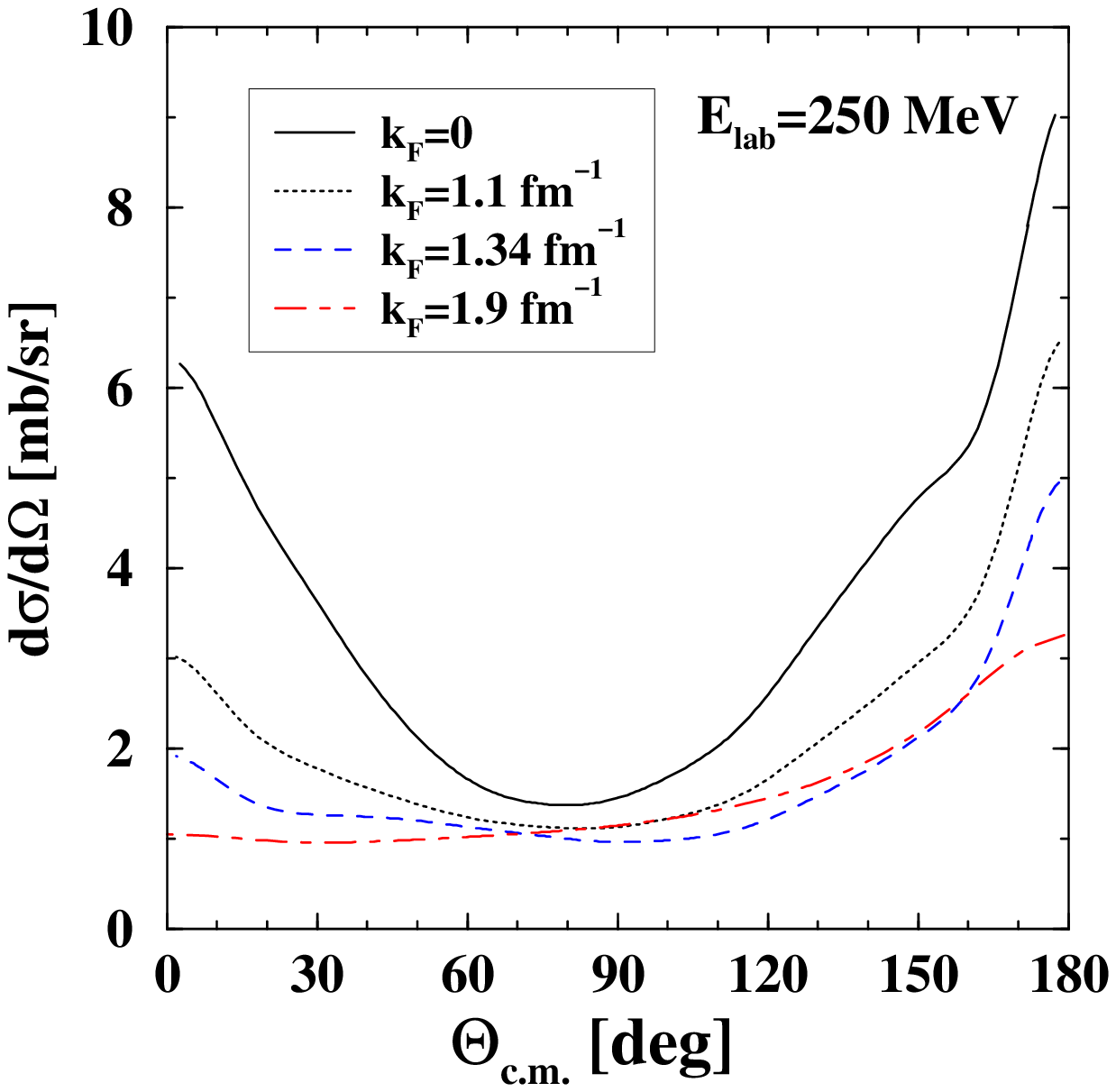}
\end{center}
\caption{Differential $np$ in-medium cross section at 
laboratory energy $E_{\rm lab}(\rho) \simeq 250$ MeV at various densities.
}
\label{fig5b}
\end{figure}

To illustrate this effect in fig.\ref{fig5b} we show the 
density dependence of the differential $np$ cross section at 
the same laboratory energy $E_{\rm lab}\simeq250$ MeV. At comparable 
energies rather than comparable c.m. momenta 
the difference in the differential cross section 
at $\rho_0$ and $3\rho_0$ is now much less pronounced. 

\begin{figure}
\begin{center}
\leavevmode
\epsfxsize = 10cm
\epsffile[100 50 460 400 ]{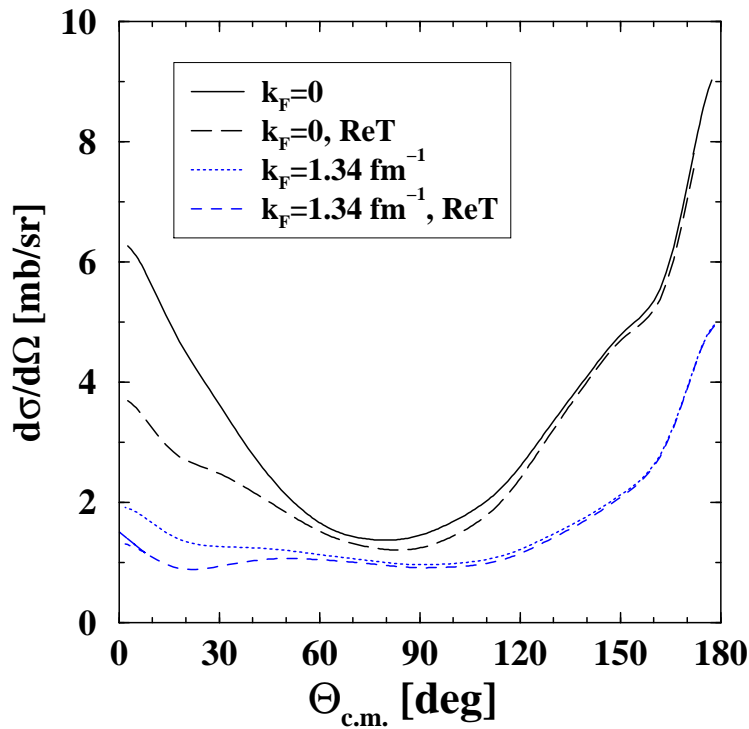}
\end{center}
\caption{Differential $np$ in-medium cross section at 
fixed relative c.m. momentum $|{\bf q}|$=342 MeV in the vacuum 
(upper curves) and at $k_F = 1.34~(\rho_0)$ (lower curves). 
The contributions from the real part of the T-matrix are shown 
separately.
}
\label{fig5}
\end{figure}

The suppression of the in-medium cross section at forward angles 
which occurs at higher densities can be understood from 
Fig.\ref{fig5} and eq. (\ref{reimt}). 
This figure illustrates the influence of the Pauli operator 
and the imaginary part 
of the T-matrix. It is seen that at $\rho_0$ the imaginary part 
of $T$ which contributes in the vacuum by about 50\%  to the 
forward scattering amplitude ($\theta =0$) is now 
strongly suppressed by the Pauli operator. This effects is 
maximal at low momenta and high densities. When ${\bf q}$ lies below 
the Fermi surface the imaginary part of $T$ vanishes completely. 
Thus, Pauli blocking in the intermediate 
states makes the cross section more isotropic. 

\begin{figure}
\begin{center}
\leavevmode
\epsfxsize = 10cm
\epsffile[100 50 460 400 ]{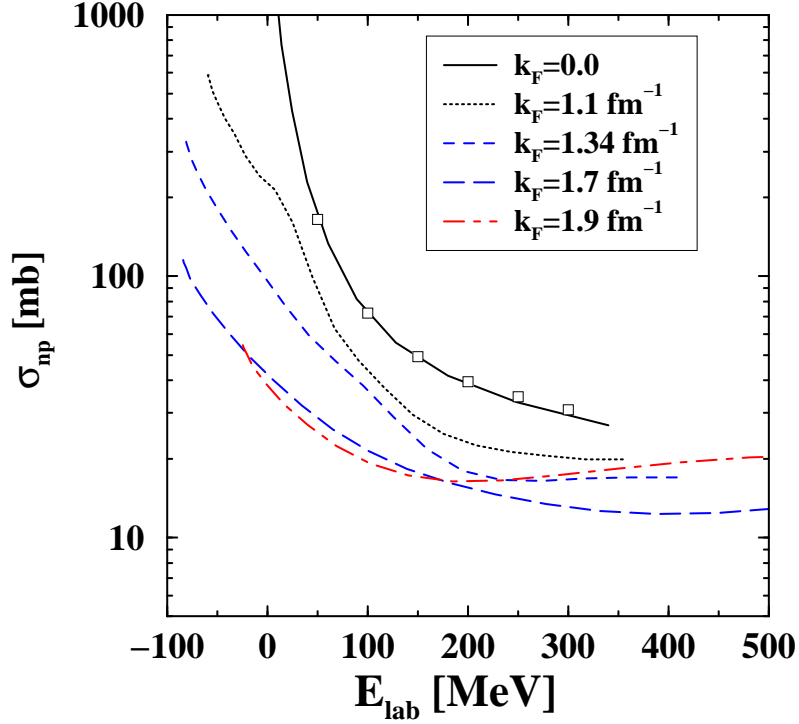}
\end{center}
\caption{Total $np$ in-medium cross section at 
various densities as a function of $E_{\rm lab}(\rho)$. In addition 
the results of \protect\cite{lima93} (squares) for the free cross section 
are shown.
}
\label{fig7}
\end{figure}

\begin{figure}
\begin{center}
\leavevmode
\epsfxsize = 10cm
\epsffile[100 50 460 400 ]{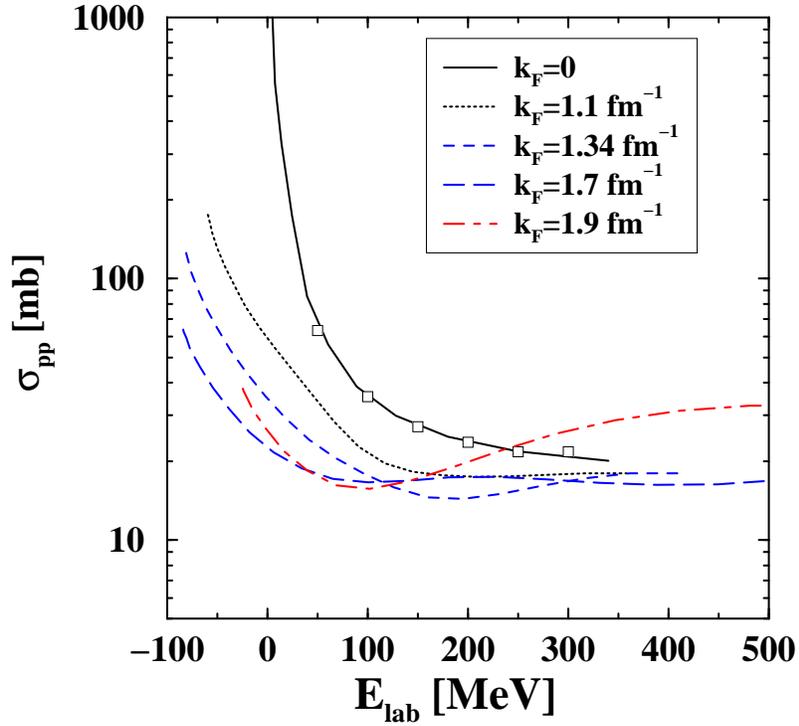}
\end{center}
\caption{Total $pp$ in-medium cross section at 
various densities as a function of $E_{\rm lab}(\rho)$.
 In addition the results of \protect\cite{lima93} (squares) 
for the free cross section are shown.
}
\label{fig8}
\end{figure}

In Figs. \ref{fig7} and \ref{fig8} we show the total in-medium 
$np$ and $pp$ cross sections in the considered density range 
$0.5\rho_0-3\rho_0$ as a function of $E_{\rm lab}$, Eq. (\ref{elab}). 
Using this quantity the scale is considerably 
stretched compared to the vacuum expression $\frac{2{\bf q}^2}{M}$ 
(used in \cite{lima93}). There are two major aspects to be noticed: 
At high energies $E_{\rm lab}\ge 200$ MeV we find good agreement 
with the previous calculations of ref. \cite{lima93}. 
For $np$ as well as $pp$ scattering the cross sections reach asymptotic 
values around 15-20 mb. At high densities 
the cross section has the tendency to rise again 
with increasing laboratory energy. 
This behaviour is even more pronounced in the 
proton-proton channel and has been observed by other 
groups as well \cite{thm87b,lima93}.    

At low energies the in-medium cross sections are considerably 
less suppressed than observed in ref. \cite{lima93}. 
One reason are our somewhat larger values for the in-medium mass, 
but this effect is not sufficient to explain the 
deviations at low energies. As illustrated by Fig.\ref{pauli.fig} 
the differences can be understood by the low density/high momentum 
approximation made in \cite{lima93} which neglects the influence 
of the Pauli operator in the optical theorem. 
Besides this point, at $0.5\rho_0$ we see a small additional 
enhancement of the $np$ cross section around an energy of 15 MeV 
which not present in the $pp$ channel. A much stronger enhancement 
of the cross section at low densities has been observed 
by Alm et al. \cite{alm94}. In the finite 
temperature approach of \cite{alm94} this critical enhancement 
is attributed to the onset of superfluidity. Crucial for such a  
superfluid state are contributions from hole-hole 
scattering in the Pauli operator which are absent in the standard 
Brueckner approach (used here). However, as discussed in 
\cite{vonderfecht} a signature of a bound pair state can appear 
at low densities even when hole-hole scattering is neglected in the 
Pauli operator. In the present calculations such an resonance like 
enhancement of the cross section is only seen in the $I=0$ 
amplitudes which correspond to the quantum numbers of the deuteron, 
i.e. the $^{3}S_1$, $^{3}D_1$ and the  $^{3}S_1$-$^{3}D_1$ transition 
channels. Therefore the low density enhancement of the $np$ cross section 
can be interpreted as a precursor of a superfluid state and supports 
the findings of \cite{vonderfecht}.

\section{half off-shell scattering}
\subsection{Matrix elements}
In the following we consider the case of half off-shell scattering. 
This means that the initial states with c.m. momenta 
$\pm {\bf q}$ are on their mass shell $E^* ( {\bf q}) = 
\sqrt{ {\bf q}^2 +M^{*2}}$. However, now the momenta of the 
final states $\pm {\bf p}$ 
can vary independently, i.e. $|{\bf p}| \not= |{\bf q}|$. 
Since the requirement of energy conservation fixes 
the final energies $p_\mu = \{p_0, \pm {\bf p}\}$ to 
$p_0 = \frac{1}{2}\sqrt{s^*} = E^* ( {\bf q})$, these states are 
off energy shell. To obtain an impression how the off-shellness 
affects the T-matrix amplitudes we consider first the matrix 
elements. In the on-shell case the cross sections follow 
from the squared matrix 
elements $|T(q,q)|^2$ by Eq. (\ref{sig1}). The squared matrix 
elements $|T(p,q)|^2= \int|T(p,q,\theta )|^2~d\Omega $ are also those  
quantities which enter directly into a generalised transport 
equation as transition amplitudes for 
off-shell scattering \cite{cassing}. In that case it makes more sense to 
speak in terms of transition amplitudes than in terms of 
cross sections. The latter ones are obtained from the 
transition amplitudes by integration over the final state 
spectral distributions. Thus, 
the cross sections depend crucially on the spectral width of the 
particles  whereas the transition amplitudes themselves 
are independent of the spectral functions. 

\begin{figure}
\begin{center}
\leavevmode
\epsfxsize = 15cm
\epsffile[100 280 560 670]{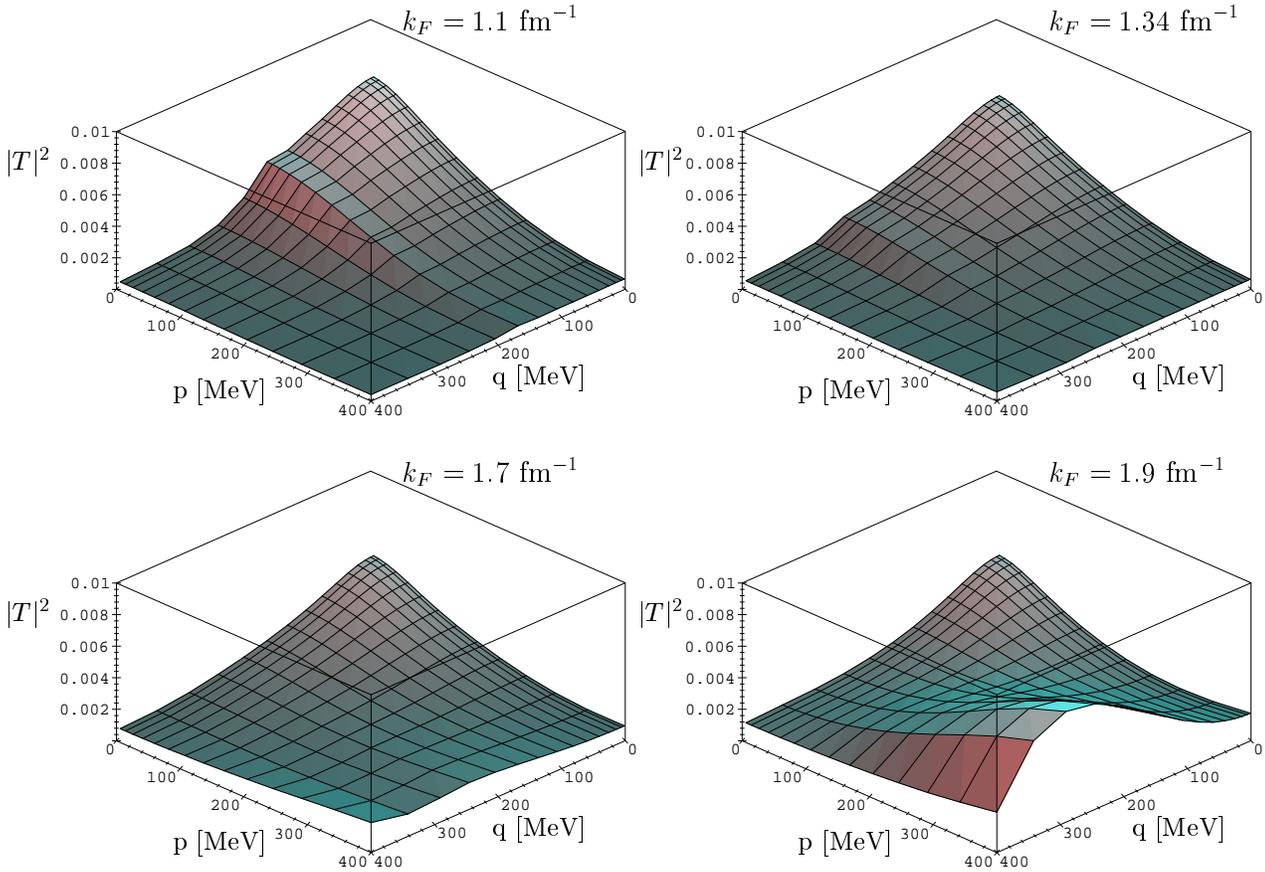}
\end{center}
\caption{Squared iso-spin averaged 
half off-shell T-matrix elements $|T|^2$ 
in [fm$^4$] at various densities. The initial states 
with relative c.m. momentum $q$ are on-shell, 
the final states with relative c.m. momentum $p$ are off-shell. 
}
\label{fig10}
\end{figure}

In Fig.\ref{fig10} we show the iso-spin averaged 
matrix elements  $|T(p,q)|^2$ for 
$NN$ scattering as a function of $p$ and $q$ at 
nuclear matter densities 0.5/1/2/3 $\rho_0$. The amplitudes $|T|^2$ are 
given in fm$^4$ and have to be multiplied by the factor 
$(2\pi)^6$ (due to our normalisation of the T-matrix) 
when they are inserted into equation 
(\ref{sig1}) to obtain cross sections. The diagonal ($p=q$) on-shell 
elements correspond to the total cross section. First of all, it is 
seen that the amplitudes  $|T|^2$ show the same 
behaviour as the corresponding (on-shell) cross sections, 
namely a general decrease with momentum and density. In the 
cross section this tendency is just enhanced by the kinematical 
factor $M^{*4}/s^*$ which decreases with density and also with 
momentum. 

The off-shellness of the final states now given by 
\beqa
\Delta\omega = E^* ({\bf q}) -  E^* ({\bf p})\simeq 
\frac{{\bf q}^2 - {\bf p}^2}{2M^*}
\eeqa
which is maximal perpendicular to the diagonal. It can be seen from 
Fig.\ref{fig10} that at low momenta 
the matrix elements are only weakly affected by 
the fact that the outgoing states are off energy-shell. This does 
not mean that the matrix elements do not change going 
away from the on-shell point. Indeed, the variation of the matrix 
elements is considerable as will become even more clear further on. 
However, the dependence of the matrix elements is nearly symmetric 
in $q$ and $p$, i.e. 
\beq
|T (p,q)|^2 \simeq |T (q,p)|^2
\label{sym1}
\eeq
 and thus not strongly affected by the off-shellness 
of the outgoing states. In particular in the low momentum region the 
the matrix elements fall off symmetrically with increasing momenta 
$p$ and/or $q$. Only at the highest density of $3\rho_0$ the asymmetry 
is rather pronounced. The off-shell variation of the outgoing 
states leads here to a resonance structure around $q=250$ MeV where 
the amplitudes increase with different 
strength in $p$ and $q$ directions. The off-shell behaviour 
is similar in the $np$ and $pp$ channels, except that the 
resonance structure at $3\rho_0$ is more pronounced in the latter 
case. The somewhat stronger off-shell dependence of the $pp$ 
scattering at high densities is also reflected in the stronger 
reduction of the corresponding total cross section shown in Figs. 
\ref{figoff1},\ref{figoff2}.

\begin{figure}
\begin{center}
\leavevmode
\epsfxsize = 15cm
\epsffile[40 60 450 450]{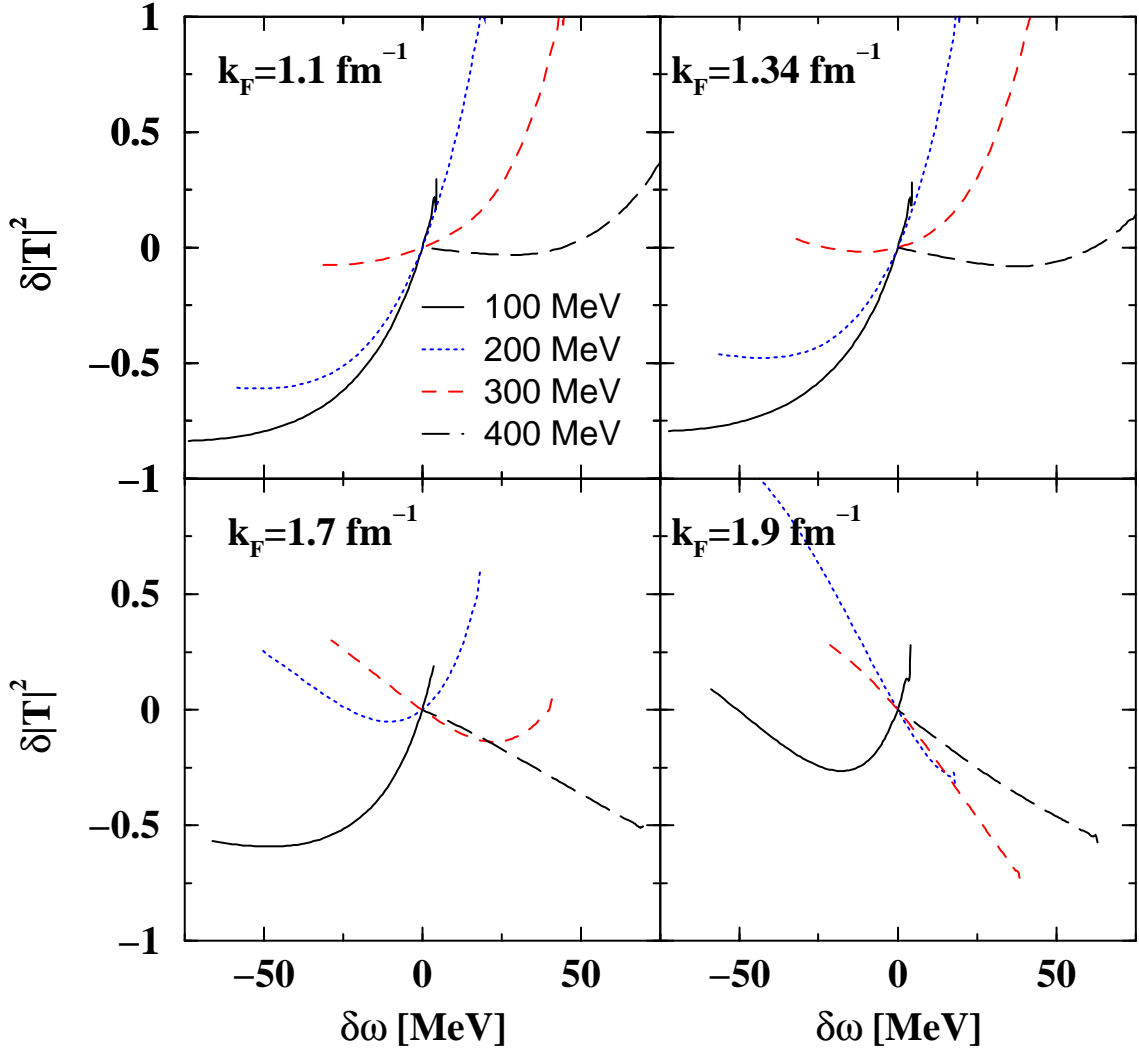}
\end{center}
\caption{Relative deviation of the isospin averaged 
T-matrix elements $|T|^2$ from their values at the on-shell 
point as a function of the 
off-shellness $\delta\omega$ of the final states. $\delta |T|^2$ is 
shown for various densities at fixed incoming relative momenta 
$q=100,~200,~300,~400$ MeV. 
}
\label{fig12}
\end{figure}

In order to obtain a more quantitative impression of the off-shell 
dependence next we investigate the deviation of the amplitudes 
$\delta |T|^2$ from their on-shell values 
as a function of the off-shellness $\delta\omega$ of 
the final states. $\delta |T|^2$ is thereby defined as the 
relative deviation, i.e.
\beq
\delta |T|^2 = \frac{|T(q,\delta\omega)|^2- |T(q,\delta\omega=0)|^2}
{ |T(q,\delta\omega=0)|^2}
\quad .
\eeq
For a better comparison of the different densities the variable 
$\delta\omega$, i.e. the energy shift of the final states with momenta 
$\pm {\bf p}$ relative to their on-shell energies, 
is scaled by the effective mass
\beq
\delta\omega = \frac{M^*}{M}(E^* ({\bf q}) - E^* ({\bf p}))
\quad . 
\eeq
The range for the variation of $\delta\omega$ is constrained by the 
kinematical limits of our calculations, 
i.e. $0\leq p,q \leq 400$ MeV. The symmetric fall off with $p$ and $q$ 
in the low momentum range which is reflected in Fig.\ref{fig10} 
implies already the following tendency: For $p < q$, i.e. 
$\delta\omega > 0$, one expects an enhancement of the amplitudes, i.e. 
$\delta |T|^2 > 0$, 
whereas for $p > q$, i.e. $\delta\omega < 0$, one expects a reduction 
of the amplitudes. For small values of $q$ this behaviour is 
clearly seen from Fig.\ref{fig12}. There the deviation of the isospin 
averaged matrix elements $\delta |T|^2$ from the on-shell point 
is shown at fixed values of the on-shell momentum $q=100,~200,~300$ and 
400 MeV. First of all we see that the variation of the amplitudes 
is pronounced. Within the range of $\delta\omega =\pm 50$ MeV 
the matrix elements can easily vary by more than $\pm 100\%$. Secondly, 
we find that the pattern are similar at moderate densities $0.5\rho_0$ 
and $\rho_0$ but become essentially different at large densities 
$\rho \ge 2\rho_0$. Thus, the off-shell behaviour reflects a strong 
medium dependence. The systematics of the different pattern is, 
however, quite complex. 

Let us first consider moderate nuclear matter densities. At low momenta 
$q$ the amplitudes show an extremely strong variation around the 
on-shell point which reflects their steep and symmetric fall off 
(see Fig.\ref{fig10}). As already mentioned, this results in a 
strong suppression at negative $\delta\omega$ 
and an equally strong enhancement at positive $\delta\omega$. In the 
high momentum region the variation of the amplitudes is much weaker 
which results in a smoother and less pronounced 
dependence on $\delta\omega$. With increasing density 
the situation changes drastically and is even reversed at 
large $q$: now the amplitudes are strongly enhanced in the 
negative  $\delta\omega$ region and reduced at positive $\delta\omega$. 
This reflects the asymmetry of $|T|^2 $ around the on-shell diagonal seen 
in Fig.\ref{fig10} in the high momentum range. 

To summarise: The half-off-shell matrix elements show a pronounced 
density dependence. At moderate densities, however, the dependence on 
the incident on-shell momenta $q$ and the final momenta $p$ which 
are off-shell is to large extent symmetric in $p$ 
and $q$. This implies that the matrix elements are not too much 
affected by the shift to off-shell energies but are mainly determined 
by the absolute values of the momentum states $p$ and $q$. As a 
consequence, the off-shell matrix elements can be approximated 
with an accuracy of about $10-30\%$ 
by the on shell points in the following way 
\beqa
|T (p,q)|^2 \simeq |T ({\bar q},{\bar q})|^2 
\quad ,\quad  {\bar q}=\sqrt{\frac{1}{2}(p^2 +q^2 )}
\label{sym2}
\eeqa
which follows from the symmetry assumption (\ref{sym1}). At large 
nuclear matter densities this symmetry is more strongly violated, 
i.e. by about $20-40\%$ at $2\rho_0$. Nevertheless, in view of the 
extremely large variation of the matrix elements with $\delta\omega$, 
ranging from almost complete suppression to an enhancement of more 
than a factor of two, the accuracy of this symmetry 
assumption is quite good in the considered $\delta\omega$ interval. 
It can be applied in a straightforward 
way, e.g. in transport calculations and requires only the 
knowledge of the on-shell matrix elements. However, at 
$3\rho_0$ the amplitudes are 
highly asymmetric in $p$ and $q$ and thus approximation (\ref{sym2}) 
does no more hold. Here an accurate description 
requires the knowledge of the exact matrix elements. 

\begin{figure}
\begin{center}
\leavevmode
\epsfxsize = 15cm
\epsffile[40 60 450 450]{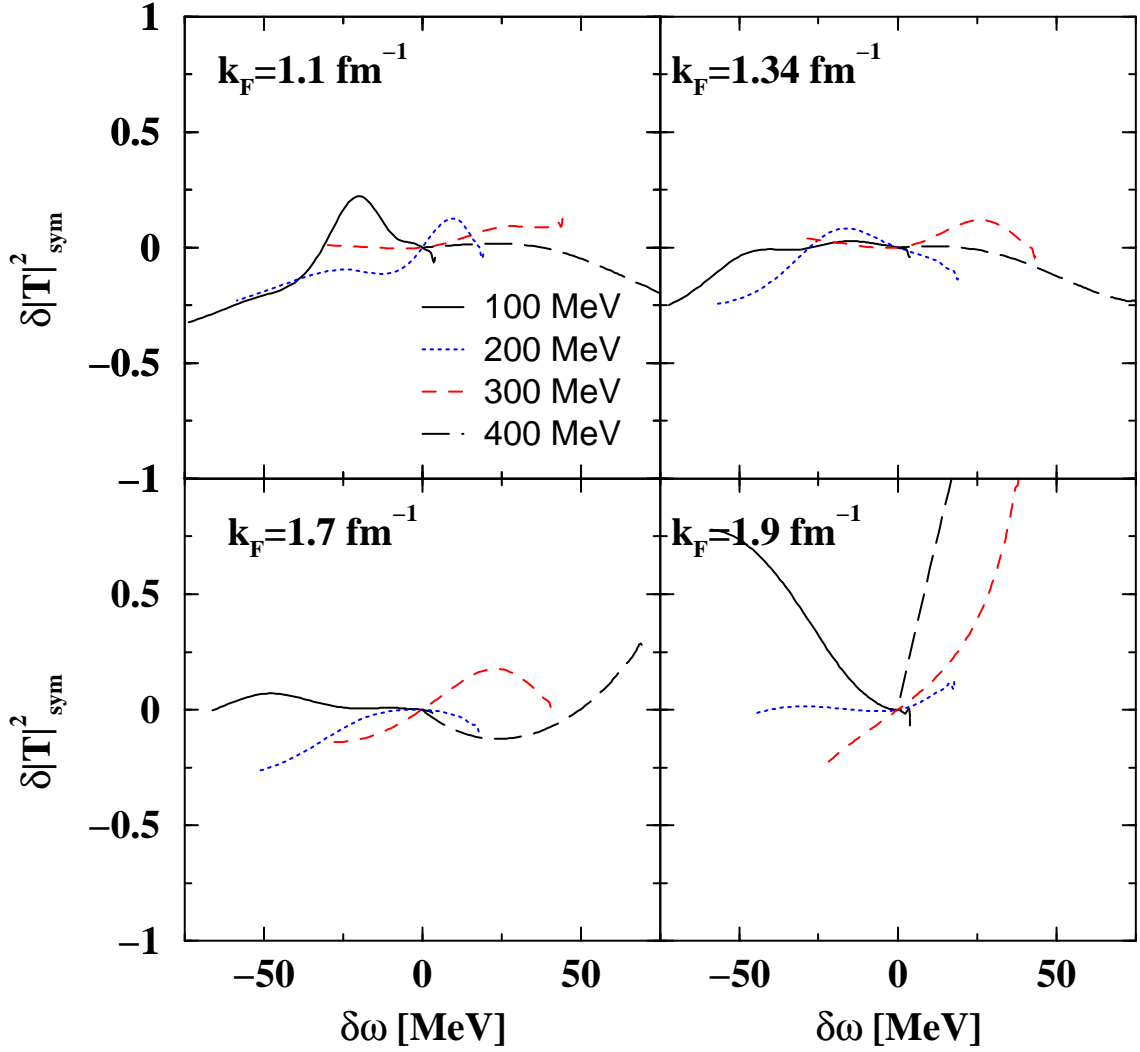}
\end{center}
\caption{Relative deviation of the isospin averaged 
approximated T-matrix elements $|T|^2$, Eq. (\protect\ref{sym2}),  
from their exact off-shell values as a function of the 
off-shellness $\delta\omega$ of the final states. $\delta |T|^2$ is 
shown for various densities at fixed incoming relative momenta of 
$q=100,~200,~300,~400$ MeV. 
}
\label{fig13}
\end{figure}

Fig.\ref{fig13} illustrates the validity of the symmetry 
assumption (\ref{sym2}) 
and the strength of the explicit dependence of the matrix elements 
on the energy shift $\delta\omega$. There the relative deviations 
\beq
\delta |T|^{2}_{\rm sym} = 
\frac{|T({\bar q},{\bar q})|^2- |T(p,q)|^2}
{ |T(p,q)|^2}
\label{sym3}
\eeq
from the exact results $|T(p,q)|^2$ are shown as functions of 
$\delta\omega$, again for fixed values of $q$.

\subsection{Cross sections}
A cross section is obtained from the transition amplitudes $|T|^2$ 
by the division through the incoming flux
\beqa
v_{12} = \frac{F^* (q_1,q_2)}{E^*({\bf q}_1) E^*({\bf q}_2)}
= \frac{\sqrt{(q_1\cdot q_2)^2 -M^{*4}}  }{E^*({\bf q}_1) E^*({\bf q}_2)}
\eeqa
and the multiplication with the final state phase space factors
\beq
d\sigma = \frac{M^{*4}}{F^*} (2\pi)^4 \delta^4 (q_1+q_2-p_1-p_2)
|T(p_1 p_2,q_1 q_2)|^2 \frac{d^4 p_1}{(2\pi)^4} A(p_1) 
\frac{d^4 p_2}{(2\pi)^4} A(p_2)
\label{sig3}
\quad .
\eeq
The $\delta^4 $-function ensures energy-momentum conservation. In the general 
case where the final states $p_1, p_2$ are off-shell, $A$ represents 
the full positive energy spectral function. 
In the quasi-particle approximation 
$A$ reduces to the on-shell condition 
\beq
A(p) = 2\pi \delta(p^{*2} - M^{*2}) 2\Theta(p^{*}_0)
\label{qpa}
\quad .
\eeq
Thus the spectral function fulfils the sum rule
\beq
\int \frac{dp_0}{(2\pi)} A(p) =  \frac{1}{E^* ({\bf p})}
\quad .
\label{sum}
\eeq
Here we choose spectral functions of a Breit-Wigner form
\beq
A(p) = 2\pi \frac{1}{\pi} 
\frac{ 2~p_0~ \Gamma}{ \left( p^{*2} - M^{*2}\right)^2 
 + \left( p_0\Gamma\right)^2 }
\quad .
\label{spec1}
\eeq
Thus $A$ satisfies the sum rule and in the zero width limit 
$\Gamma \longmapsto 0$ the quasiparticle approximation (\ref{qpa}) 
is recovered. In the two-particle c.m. frame the integral (\ref{sig3}) can 
easily be evaluated. The total cross section is then given by
\beqa
d\sigma = &&\frac{2 M^{*4}}{ \pi^4 \sqrt{s^* (s^* - 4M^{*2})}} 
d\Omega \int p^{2} dp~ |T(p,q,\theta )|^2 
\nonumber \\
&& \int  
\frac{dp_0 ~ (2E^*({\bf q}) - p_0)~ p_0~ \Gamma^2 }{ 
\left[ \left( ( 2E^*({\bf q}) - p_0)^2 - (p^2 + M^{*2}) \right)^2 
 + ( 2E^*({\bf q}) - p_0)^2 ~\Gamma^2 \right]
\left[ (p_{0}^2 - p^2  - M^{*2} )^2 + p_{0}^2 ~\Gamma^2\right] }
\label{spec2}   
\eeqa
which reduces to expression (\ref{sig1}) in the zero width limit. 
As discussed in \cite{btm90} the width $\Gamma$ is determined by the 
imaginary part of the self-energy 
\beq
\Gamma = - \frac{M^*}{E^* ({\bf p})}Im\Sigma_s (\rho,{\bf p}) + 
Im\Sigma_0 (\rho,{\bf p})
\label{gam1}
\eeq
and depends on density and momentum. In the present approach 
$\Gamma$ follows from the imaginary part of the T-matrix. It can, 
however, only serve as an estimate for the full particle width inside 
the medium. As discussed in Sec. II the Bethe-Salpeter equation is 
solved in the quasi-particle approximation. For the determination 
of the $T$ matrix this treatment is justified since the 
condition $\Gamma \ll E_{\rm s.p.} = E^* - \Sigma_0$ is readily 
fulfilled. However, since the standard Brueckner-Hartree-Fock 
approach accounts only for the particle-particle correlations 
of the Brueckner hole-line expansion $Im\Sigma$ vanishes for momenta 
below the Fermi surface due to Pauli blocking. Long-range correlations 
which are usually treated in RPA-type approaches by a ladder summation 
in the particle-hole channel contribute to the spectral 
width \cite{muether2000}, but are not taken into account 
in the standard Brueckner-Hartree-Fock. RPA correlations 
lead further to a depletion just below the Fermi surface and an 
occupation of states above the Fermi surface. Thus, the DBHF results 
for $\Gamma$ do not represent the full width 
but give a reliable estimate only at momenta well above the Fermi 
surface where particle-particle correlations can be regarded 
as the the dominant contributions to $\Gamma$. 
In the considered density and energy range 
$\Gamma$ ranges from about 10 MeV at $0.5\rho_0$ to 
more than 40 MeV at $3\rho_0$ as the outcome of the 
present DBHF calculations \cite{boelting99}.

\begin{figure}
\begin{center}
\leavevmode
\epsfxsize = 15cm
\epsffile[40 60 450 450]{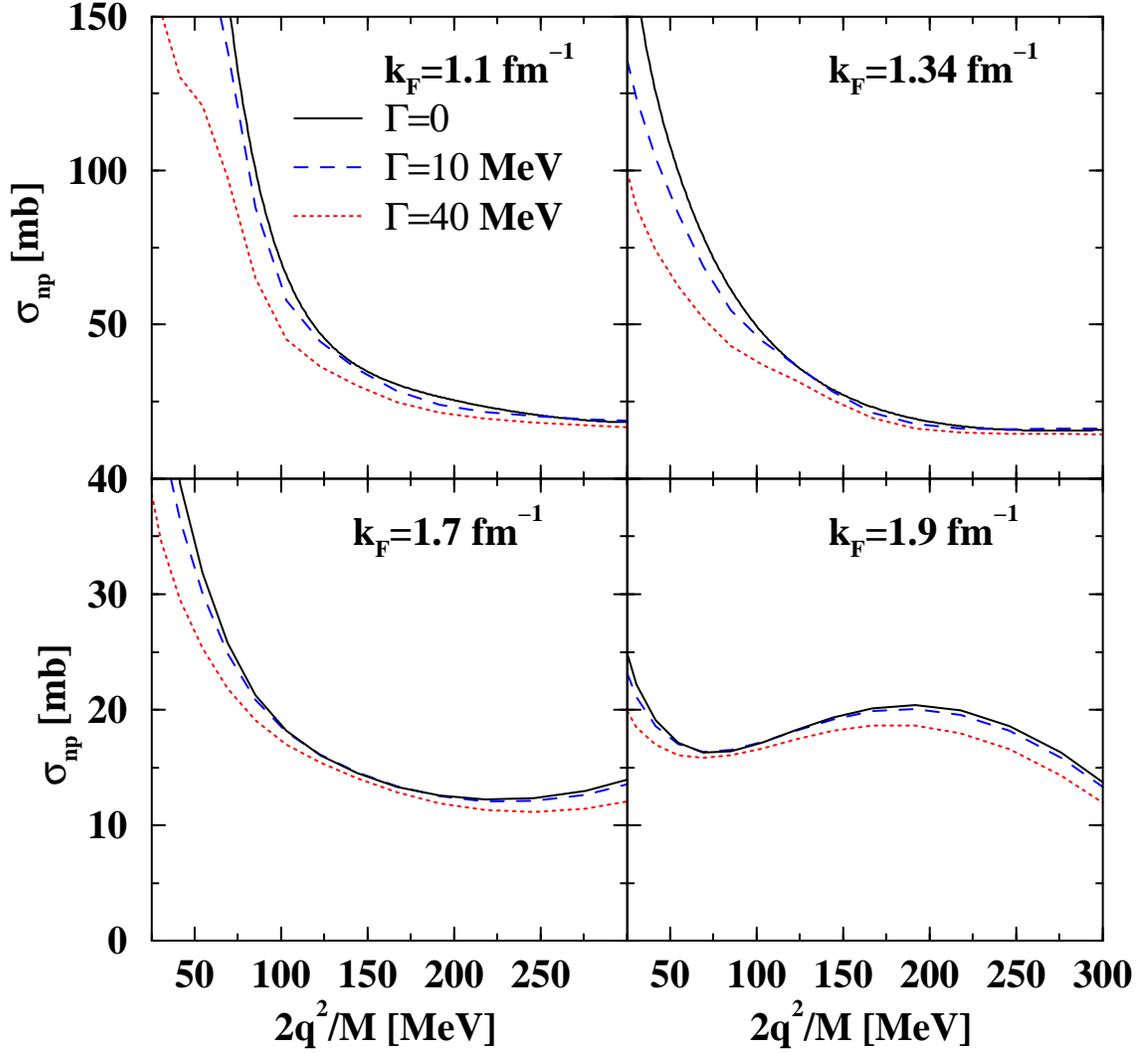}
\end{center}
\caption{Total $np$ cross section at various densities as a function of 
$2{\bf q}^2/M$. The quasi-particle approximation ($\Gamma = 0$) is 
compared to the case where the final states have a finite spectral 
width of $\Gamma = 10,~40$ MeV.
}
\label{figoff1}
\end{figure}

\begin{figure}
\begin{center}
\leavevmode
\epsfxsize = 15cm
\epsffile[40 60 450 450]{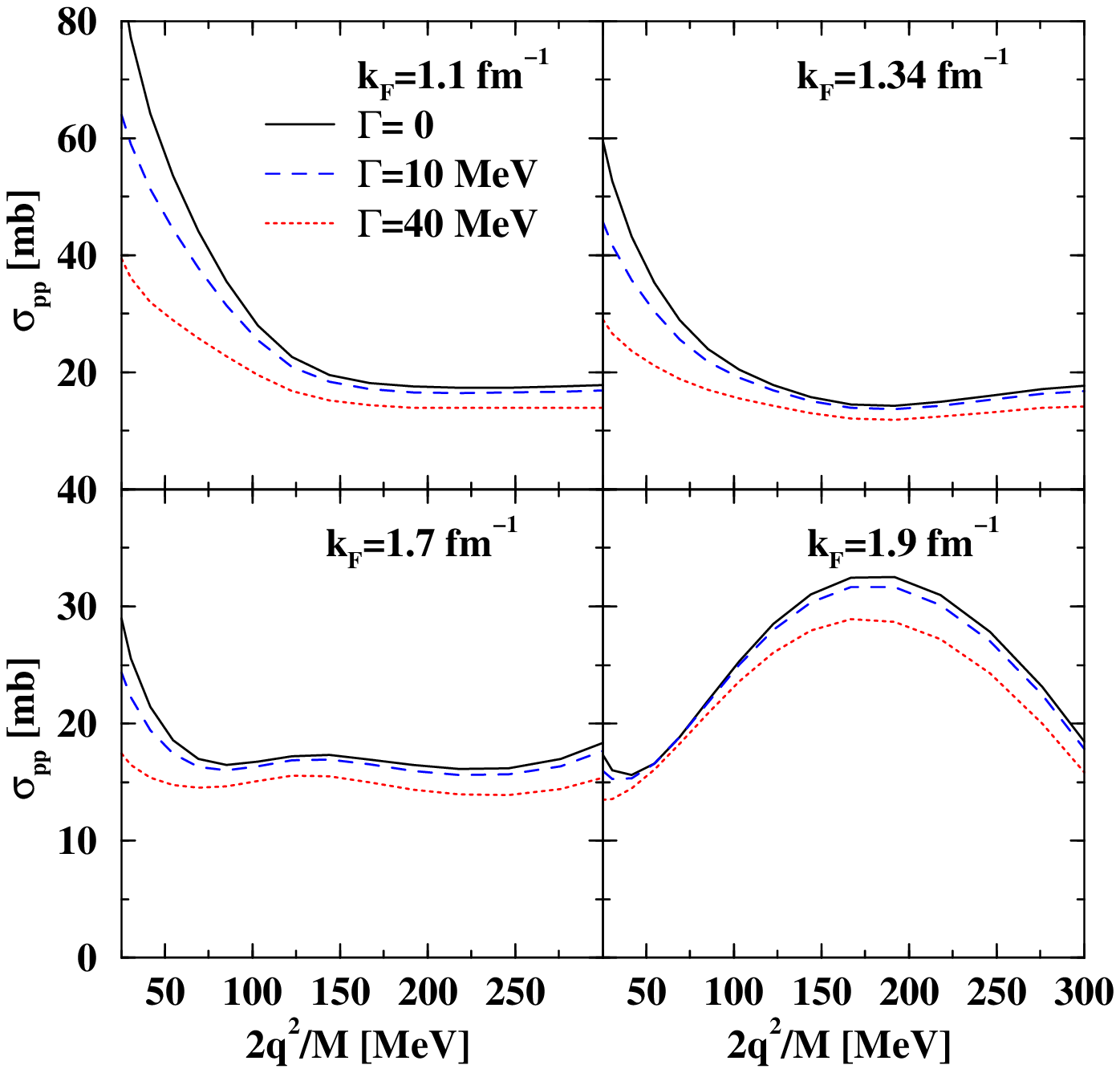}
\end{center}
\caption{Total $pp$ cross section at various densities as a function of 
$2{\bf q}^2/M$. The quasi-particle approximation ($\Gamma = 0$) is 
compared to the case where the final states have a finite spectral 
with of $\Gamma = 10,~40$ MeV.
}
\label{figoff2}
\end{figure}

Thus, the cross sections shown in Figs. \ref{figoff1} ($np$) and 
\ref{figoff2} ($pp$) 
give an estimate for the off-shell dependence of the 
total cross section. For this purpose we choose two typical values 
for $\Gamma$ which cover the range of the spectral width in nuclear 
matter as it is predicted by DBHF calculations, i.e. $\Gamma = 10$ MeV 
and 40 MeV. The present results are obtained with a constant, i.e. 
momentum independent $\Gamma$. For a better comparison of the various 
densities the cross sections are shown as a function of $2{\bf q}^2/M$ 
which corresponds in the vacuum to the laboratory energy. 
It can be seen from Figs. \ref{figoff1}, \ref{figoff2} that 
the off-shell dependence of the total cross section is 
moderate. The averaging over the Breit-Wigner distributions 
leads to significantly smaller off-shell 
effects than are seen in the scattering amplitudes. 
Compared to the on-shell value the cross sections are 
generally reduced, in the case of $\Gamma = 40$ MeV by about 15-20\%. 
At small nuclear densities this reduction is most pronounced at low 
momenta whereas but at higher momenta there is no sizable effect. 
At large densities the reduction is more pronounced at high momenta. Only 
for large values of $\Gamma\simeq 100$ MeV (not shown here) 
a reduction of the cross section of more than 50\% can be reached. 
Differences in the off-shell structure of the NN interaction, 
e.g. using BONN C, do not lead to essentially different results. 
This is consistent with the observations made in Ref. \cite{lima93} 
where it has been demonstrated that the on-shell cross sections 
are not much affected by the use of different parameterisations 
like BONN B or C.

\section{Summary}
In the present work we investigated the in-medium nucleon-nucleon 
cross section within the relativistic (Dirac) Brueckner-Hartree-Fock 
approach. We considered both, on-shell scattering and the more general 
case where the final momenta are allowed to be off energy shell. The 
in-medium cross sections can serve as input for transport 
calculation of heavy ion collisions. Information on the off-shell 
dependence of the scattering amplitudes is required when one intends 
to go beyond the quasi-particle approximation in order to obtain a 
more realistic description of transport phenomena for particles with 
finite width. In a dense hadronic environment created, e.g.,  in 
heavy ion collisions also 'stable particles' like nucleons acquire 
a finite spectral width. Throughout this work we applied the BONN A 
potential as the nucleon-nucleon interaction since this potential 
yields the most reasonable saturation properties for nuclear matter. 

Concerning the on-shell scattering we find a qualitative agreement 
with previous investigations of Li and Machleidt \cite{lima93}, 
however, less reduced in-medium cross sections at 
low energies. The reason therefore lies in the fact that 
we account for modifications of the optical theorem due 
to the presence of the medium which have been neglected in the 
approximation made in \cite{lima93}. In the $np$ cross section 
an additional low density enhancement appears which can be interpreted 
as the precursor of a superfluid state. 
The present approach was then extended 
to the case where incoming and outgoing momenta of the scattered 
nucleons can vary independently and - due to energy-momentum conservation - 
the final states are off energy shell. The resulting T-matrix 
elements or transition amplitudes show a strong variation around the 
on-shell point. The shape of the transition amplitudes depends, however, 
mainly in a symmetric way on the incoming/outgoing momenta. The 
fact that the final states are off-shell, the incoming ones 
on-shell, plays thereby a minor role. This allows to approximate the 
half-off-shell matrix elements in a suitable way by their on-shell 
values. Such an approximation works well at moderate nuclear matter 
densities but at large densities the precise knowledge of the amplitudes 
is required. In a conservative estimate finite width 
effects lead to a reduction of the total cross sections 
by about 20\% compared to the on-shell scattering. Off-shell 
effects are more pronounced at larger nuclear densities. 

        

\begin{thebibliography}{99}

\bibitem{ms59} 
P.C. Martin and J. Schwinger, Phys. Rev. {\bf 115}, 1342 (1959).

\bibitem{glw80} 
S.R. de Groot, W.A. van Leeuwen, C.G. van Weert, 
{\it Relativistic Kinetic Theory}, 
(North Holland, Amsterdam, 1980).

\bibitem{btm90}
   W. Botermans, R. Malfliet, 
   Phys. Reports {\bf 198}, 115 (1990).


\bibitem{jae92}
J. Jaenicke, J. Aichelin, N. Ohtsuka, R. Linden, 
A. Faessler, Nucl. Phys. {\bf A536}, 201 (1992).

\bibitem{gmat2}
D.T.~Khoa, N.~Ohtsuka, M.A.~Matin, A.~Faessler, S.W.~Huang, E.~Lehmann, 
R.K.~Puri, Nucl. Phys. {\bf A548}, 102 (1992).

\bibitem{zheng}
Y.-M. Zheng, C.M. Ko, B.-A. Li, and Bin Zhang, Phys. Rev. Lett. 
{\bf 83}, 2534 (1999).


        \bibitem{hs87}
        C. J. Horowitz, B. D. Serot, Nucl. Phys. {\bf A464}, 613 (1987).
        
        \bibitem{thm87}
        B. ter Haar, R. Malfliet, 
        Phys. Rep. {\bf 149}, 207 (1987).

        \bibitem{nupp89}
        C. Nuppenau, Y. J. Lee, A. D. MacKellar,
        Nucl. Phys. {\bf A504}, 839 (1989).  

\bibitem{bm90}
R. Brockmann, R. Machleidt, Phys. Rev. {\bf C42}, 1965  (1990);\\
{\it ibid.} nucl-th/9612994.


\bibitem{sehn97}
L. Sehn, C. Fuchs and A. Faessler,
Phys. Rev. {\bf C56}, 216 (1997).

\bibitem{fuchs98}
C. Fuchs, T. Waindzoch, A. Faessler and D.S. Kosov,
Phys. Rev. {\bf C58}, 2022 (1998).

\bibitem{boelting99}
T. Gross-Boelting, C. Fuchs, and A. Faessler,
Nucl. Phys. {\bf A648}, 105 (1999).

        \bibitem{bonn}
        R. Machleidt, 
        Advances in  Nuclear Physics,  {\bf 19}, 189,
        eds. J. W. Negele, E. Vogt, (Plenum, N.Y., 1989).

        \bibitem{sw86}
        B. D. Serot, J. D. Walecka, 
        Advances in  Nuclear Physics, {\bf 16}, 1,
        eds. J. W. Negele, E. Vogt, (Plenum, N.Y., 1986)



\bibitem{schmidt90}
M. Schmidt, G. R\"opke, and H. Schulz, Ann. Phys. (N.Y.) 
{\bf 202}, 57 (1990).

\bibitem{alm94}
T. Alm, G. R\"opke, and M. Schmidt, Phys. Rev. {\bf C50}, 31 (1994).

        \bibitem{thm87b} 
        B. ter Haar, R. Malfliet, 
        Phys. Rev. {\bf C36}, 1611 (1987).

        \bibitem{malfliet88}
        R. Malfliet,
        Prog. Part. Nucl. Phys. {\bf 21}, 207 (1988).

\bibitem{cross}
   C. Fuchs, L. Sehn, H.H. Wolter, Nucl. Phys. {\bf A601}, 505 (1996).

\bibitem{lima93}
G.Q. Li and R. Machleidt, 
Phys. Rev. {\bf C48}, 1702 (1993); 
{\it ibid} {\bf C49}, 566 (1994). 

\bibitem{knoll98}
Yu. Ivanov, J. Knoll, D.N. Voskresensky, 
Nucl. Phys. {\bf A657}, 413 (1999).

\bibitem{henning95}
P. Henning, Phys. Reports {\bf 253}, 235 (1995).

\bibitem{fauser95}
R. Fauser, H.H. Wolter, Nucl. Phys. {\bf A584}, 604 (1995).

\bibitem{morawetz99}
K. Morawetz, H.S. K\"ohler, Eur. Phys. J. {\bf A4}, 291 (1999). 

\bibitem{leupold}
S. Leupold, Nucl. Phys. {\bf A672}, 475 (2000); \\
J. Lehr, M. Effenberger, H. Lenske, S. Leupold, U. Mosel, 
Phys. Lett. {\bf B483}, 324 (2000).

\bibitem{cassing}
W. Cassing, S. Juchem, Nucl. Phys. {\bf A665}, 377 (2000); 
{\it ibid} Nucl. Phys. {\bf A672} (2000) 417; 
{\it ibid} Nucl. Phys. {\bf A677}, 445 (2000).

\bibitem{morawetz}
K. Morawetz, V. Spicka, P. Lipavsky, G. Kortemeyer, 
Ch. Kurts, R. Nebauer, Phys. Rev. Lett,  {\bf 82}, 3767 (1999); 
K. Morawetz, P. Lipavsky, V. Spicka, N.-H. Kwong,  Phys. Rev. 
 {\bf 59}, 3052 (1999).

        \bibitem{15}
        G. E. Brown, A. D. Jackson,
        {\em The Nucleon Nucleon Interaction}
        (North Holland, Amsterdam, 1976).

        \bibitem{erkelenz} 
        K. Erkelenz, Phys. Reports {\bf 13}, 191 (1974).

        \bibitem{11} 
        M. I. Haftel, F. Tabakin, Nucl. Phys. {\bf A158}, 1 (1970). 

        \bibitem{trefz}
        M. Trefz, A. Faessler, W. H. Dickhoff, Nucl. Phys. {\bf A443}, 
        499 (1985).

        \bibitem{16}
        M. Rose, Elementary Theory of Angular Momentum (Wiley, N.Y., 1957).

\bibitem{shepard74}
P. F. Shepard {\it et al.}, Phys. Rev. {\bf D10}, 2735 (1974).

\bibitem{vonderfecht}
B. E. Vonderfecht, C. C. Gearhart, and W. H. Dickhoff, A. Polls, and 
A. Ramos, Phys. Lett. B {\bf 253}, 1 (1991).

\bibitem{muether2000}
H. M\"uther, A. Polls, Prog. Part. Nucl. Phys. {\bf 45}, 243 (2000).

\end{thebibliography}
\end{document}